\documentstyle[12pt,worldsci]{article}

\def\PL #1 #2 #3 {Phys. Lett.~{\bf#1} (#2) #3}
\def\NP #1 #2 #3 {Nucl. Phys.~{\bf#1} (#2) #3}
\def\ZP #1 #2 #3 {Z.~Phys.~{\bf#1} (#2) #3}
\def\PR #1 #2 #3 {Phys. Rev.~{\bf#1} (#2) #3}
\def\PRD #1 #2 #3 {Phys. Rev.~D {\bf#1} (#2) #3}
\def\PP #1 #2 #3 {Phys. Rep.~{\bf#1} (#2) #3}
\def\PRL #1 #2 #3 {Phys. Rev.~Lett.~{\bf#1} (#2) #3}

\newcounter{eqletter}

\begin{document}
\baselineskip14pt
\begin{flushright}
FSU--HEP--930903\\
MAD/PH/789\\
September 1993\\
\end{flushright}
\title{Measuring Three Vector Boson Couplings in $qq\to qqW$ at the SSC
\footnote{To appear in the Proceedings of the Workshop {\it ``Physics at
Current Accelerators and the Supercollider''}, Argonne National
Laboratory, June~2 --~5, 1993.}
}
\author{U.~Baur$^1$ and D.~Zeppenfeld$^2$\\
\it $^1$Department of Physics, Florida State University, Tallahassee, FL
32306,
USA\\
$^2$Department of Physics, University of Wisconsin, Madison, WI 53706, USA}
\maketitle

\begin{center} ABSTRACT\\ [.1in]
\parbox{13cm}{\small
We investigate the electroweak process $qq\to qqW$, {\it i.e.} $W$
production via
$W\gamma$ and $WZ$ fusion, as a probe for nonstandard $WW\gamma$ and $WWZ$
vertices at
hadron supercolliders. Triggering on events with one very forward and one very
backward jet while requiring the $W$ decay lepton to be central, strongly
enhances the triple gauge boson vertex contribution and suppresses
backgrounds from QCD $Wjj$ events and $t\bar tj$ production below the signal
level. At the SSC, the process is sensitive to $WWV$, $V=\gamma\, ,Z$
couplings
$\kappa_V -1$, $\lambda_V$, and $g_1^Z$ in the $0.03\dots 0.1$ range. }
\end{center}

\section{Introduction}

The study of electroweak processes is one of the main tasks of experiments at
the SSC or the LHC. In order to probe the interactions in the bosonic sector
of the Standard Model (SM) as completely as possible one would like to
have available a large
number of basic processes and correspondingly a large number of observables.
This includes the search for the Higgs boson in both gluon fusion and weak
boson fusion reactions in order to measure the Higgs couplings to known
particles. Electroweak boson pair production ($q\bar q\to W\gamma,\; WZ$
and $W^+W^-$) probes the non-abelian $WW\gamma$ and $WWZ$ couplings.
Quartic weak
boson couplings are probed in elastic weak boson scattering and multiple weak
boson production.

Here we report on investigations of a process which will be complementary to
weak boson pair production in the measurement of the $WW\gamma$ and $WWZ$
triple gauge boson vertices (TGV's), namely single $W$ production via the
electroweak process $qq\to qqW$ (to be called ``signal'' process in the
following). Representative Feynman graphs are shown in Fig.~1.
\begin{figure}[t]
\vglue1.0in
\includegraphics{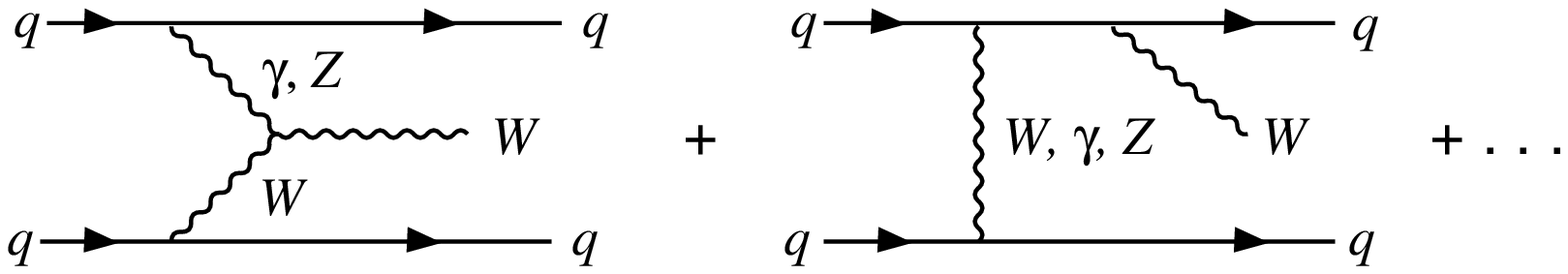}
\begin{center}
{\small Fig.~1: Feynman graphs for the $qq\to qqW$ signal. }
\end{center}
\vskip -0.5cm
\end{figure}

It is via the first graph, the $W\gamma$ or $WZ$ fusion contribution, that
this process is sensitive to deviations of the TGV's from their SM
predictions.  While $W\gamma$, $WZ$, or $W^+W^-$ production at hadron or
$e^+e^-$ colliders probe the TGV's for time-like momenta of all interacting
electroweak bosons, the   signal process measures these TGV's for space-like
momentum transfer of two of the three gauge bosons and hence is complementary
to the pair production processes.

In the following we use the standard parameterization of the TGV's in terms
of $C$ and $P$ conserving anomalous couplings $g_1^V,\; \kappa_V$ and
$\lambda_V$
to quantify the sensitivity of the signal. These couplings can be defined by
the effective Lagrangian\cite{HPZH}
\begin{equation}
i{\cal L}_{eff}^{WWV} = g_{WWV}\left( g_1^V(
W^{\dagger}_{\mu\nu}W^{\mu}-W^{\dagger\, \mu}W_{\mu\nu})V^{\nu} +
\kappa_V\,  W^{\dagger}_{\mu}W_{\nu}V^{\mu\nu} + {\lambda_V\over m_W^2}\,
W^{\dagger}_{\rho\mu}{W^{\mu}}_{\nu}V^{\nu\rho}\right) . \label{LeffWWV}
\end{equation}
Here the overall coupling constants are defined as $g_{WW\gamma}=e$ and
$g_{WWZ}= e \cot\theta_W$ where $\theta_W$ is the Weinberg angle.
Within the SM, the couplings are given by
$g_1^Z = g_1^\gamma = \kappa_Z = \kappa_\gamma = 1$, and $\lambda_Z =
\lambda_\gamma = 0$. $g_1^\gamma$ is just the electric charge of the $W$ and
hence fixed to 1 by electromagnetic gauge invariance. This leaves five
couplings which need to be determined experimentally. Deviations of these
TGV's from their SM values lead to amplitudes, e.g. for $W$ pair production,
which grow with energy, eventually violating partial wave
unitarity\cite{joglekar}.
Hence, anomalous TGV's must actually be form-factors, decreasing at large
momentum transfer, a consideration which is essential at hadron supercolliders
with their large available energy. Because of these form-factor effects the
TGV's which can be measured at space-like momentum transfer in $qq\to qqW$ may
in fact be different from the ones probed in vector boson pair production at
space-like momenta.

\section{Signal and Background Calculation}

The signal and all background cross sections were calculated using parton level
Monte Carlo programs. For the signal the program evaluates the tree level
cross sections for the process $q_1q_2\to q_3q_4\ell\nu$, $\ell=e,\mu$ and
all relevant
crossing related processes. A more detailed discussion of the signal
calculation is given in Ref.~3; our code is identical to the one used
there. For the quark and gluon structure functions inside the proton we use
set HMRS(B) of Harriman {\it et al}.\ \cite{HMRSB} for both signal and
background processes.

Anticipating large backgrounds, all the features of the final state $Wjj$
system need to be exploited for background suppression. Hence, we only
consider the signal in the case when both final state (anti)quarks have
transverse momenta larger than 40 GeV, allowing their identification as
hadronic jets. Even though the resulting total cross section is sizable,
$\sigma_{\rm sig} (pp\to W^\pm jj) = 203\; {\rm pb}$ at the SSC, one still
needs to fight much larger backgrounds. They mainly arise from Drell-Yan
production of $W$'s and from $t\bar t$ production with subsequent $t\to Wb$
decay.

The dominant background source is Drell-Yan production of $W$'s with an
expected cross section of 100 to 300 nb\cite{kuijf} at the SSC, 3 orders of
magnitude larger than the $Wjj$ signal. The background cross sections
including the additional two final state jets are calculated via the
${\cal O}(\alpha^2_s)$ real emission corrections to the Drell-Yan process. We
use a parton level Monte Carlo program based on the work of Ref.~6
to model this ``QCD $Wjj$'' background. The scales of the parton distribution
functions and of the strong coupling constant $\alpha_s(Q^2)$ are chosen to
be the transverse energy of the produced $W$. One may wonder whether double
parton scattering (DPS) is an important source of background events. Here one
pair of partons would yield a $Wj$ final state and the second parton pair
gives rise to
a dijet system, supplying the second required jet. This question has been
analyzed in Ref.~3 with the result that DPS is smaller than the QCD
$Wjj$ background by roughly one order of magnitude. Since we use acceptance
cuts similar to the ones of Ref.~3 we expect the same relative
suppression and we shall neglect the DPS background in the following.

$t\bar t$ production with subsequent $t\to Wb$ decay is another prominent
source for $W$'s. For a top-quark mass of
$m_t=140$~GeV the production cross section is $\sigma(pp\to t\bar t
X)\approx 15$~nb~\cite{tt} and hence about two orders of magnitude larger than
the $Wjj$ signal. A characteristic feature of the $W$ signal is the presence
of two energetic forward jets. In Ref.~8 it was shown in connection
with single forward jet tagging that the dominant source of forward jets in
$t\bar t$ events arises from QCD radiation, {\it i.e.}\ the additional parton
in $t\bar t j$ events, and not from the top decay products. Hence we model the
top background with a tree level Monte Carlo program based on the matrix
elements of Ref.~9 for the
processes $pp \to t\bar t j\to W^+b\, W^-\bar b j$. A phase space distribution
is assumed for the subsequent $t\rightarrow Wb$ and
$W\rightarrow\ell\nu,\, q\bar q'$ decays. While the top background
analysis is performed with a top quark
mass of 140~GeV, we have checked that a mass as light as 110~GeV does not
change the top background level qualitatively.

\section{Acceptance Cuts and Background Suppression}

In order to gain good sensitivity to the TGV we need to identify the phase
space region in which the electroweak boson fusion graph of Fig.~1 is
important. Clearly this graph is enhanced at small $Q^2$ of the incident
$\gamma,Z$ and $W$, of order $m_W^2$ or less. These virtualities are much
smaller than the typical lab frame energies of the scattering quarks which,
therefore, emerge at very small angles. Hence we want to tag events with one
very forward and one very backward jet (arising from the spectator quark
jets), while the lepton originating from the $W\to \ell\nu$ decay is to
be expected in the central region.

A search algorithm for the signal events has been outlined in Ref.~3,
emphasizing the ``rapidity gap'' characteristics of the signal. We
closely follow this approach for the lepton and jet acceptance cuts but do
not require low hadronic activity in the central region. Events are triggered
by a charged lepton of transverse momentum
\begin{equation}\label{ptl}
p_{T\ell} > 20\; {\rm GeV}\; ,
\end{equation}
and we require missing transverse momentum in excess of 50 GeV as a signature
for $W$ leptonic decays. On either side of the charged lepton (with respect
to pseudorapidity) one then searches for the first hadronic jet with
\begin{equation}\label{ptjetaj}
p_{Tj} > 40\; {\rm GeV}\; , \quad    |\eta_j|<5\; ,
\end{equation}
which will be called tagging jets and represent the two spectator quarks in
our signal calculation. Leptons and jets are required to be well separated
\begin{equation}\label{deltaR}
R_{jj} = (\Delta\eta_{jj}^2+\Delta\phi_{jj}^2)^{1\over 2} > 0.7\; ,  \quad
R_{\ell j} = (\Delta\eta_{\ell j}^2+\Delta\phi_{\ell j}^2)^{1\over 2} > 0.7\;
{}.
\end{equation}
The forward-backward nature of the two tagging jets is then taken into account
by requiring
\begin{equation}\label{etaj}
-5<\eta_{j_1}<-2.5\;, \quad  2.5<\eta_{j_2}<5\; .
\end{equation}
Notice that this implies the existence of a central ``rapidity gap'', at
least 5 units wide in pseudorapidity, which contains the charged lepton but
no jets with $p_T>40$~GeV.

The above requirements leave a QCD $Wjj$ background which is about a factor
six larger than the remaining signal. However, the background is dominated by
$W$-bremsstrahlung off initial or final state quarks, a class of events which
is also present in the electroweak signal and obscures the contribution from
the electroweak boson fusion graph. $W$-bremsstrahlung and electroweak boson
fusion lead to drastically different lepton pseudorapidity distributions for
the signal and the QCD background.
\vglue3.4in
\includegraphics{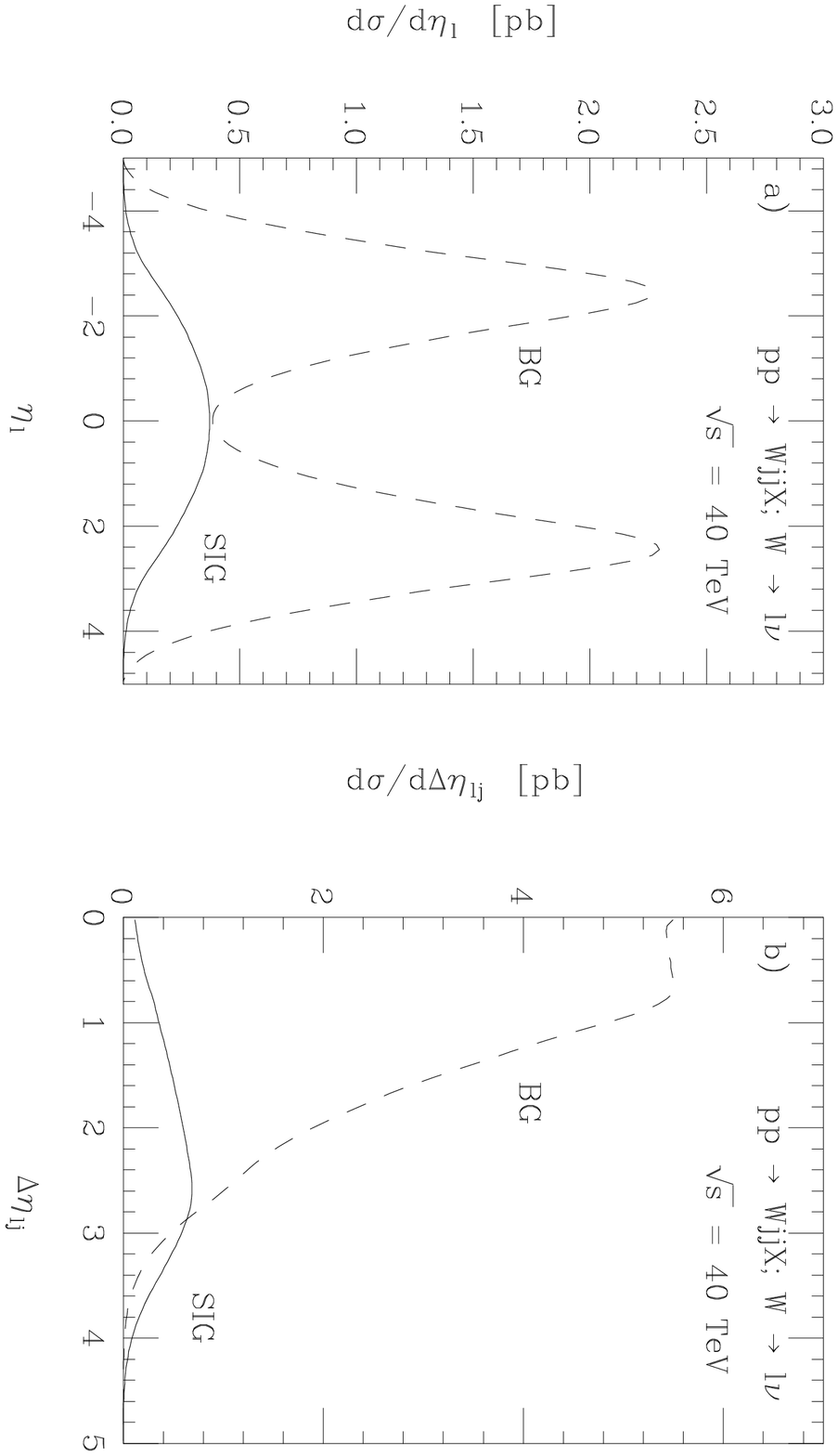}
\begin{center}
\parbox{13cm}{\small\baselineskip12pt
 Fig.~2: a) Pseudorapidity distribution $d\sigma/d\eta_{\ell}$ of the charged
decay lepton in $Wjj,\, W\to\ell\nu$ events at the SSC.
b) Minimum separation in pseudorapidity of the decay lepton from the two
tagging jets. Distributions are given for the signal (solid lines) and the
QCD background (dashed lines). }
\end{center}

Shown in Fig.~2a is the $\eta_\ell$ distribution of the charged lepton.
$W$-bremsstrahlung emits the $W$'s and hence also the decay leptons either
close to the beam pipe or close to the tagging jets, which are also required
to be very forward. As a result central lepton rapidities are rare for the
QCD background if the jet tagging cuts of Eq.~5 are imposed,
while weak boson fusion favors central leptons. Closely
related are the distributions of Fig.~2b which show the smallest separation
of the decay lepton from the two tagging jets. Again the dominance of
$W$-bremsstrahlung in the QCD background is evident. It is strongly suppressed
by requiring
\begin{equation}\label{separationlepton}
|\eta_\ell|<1.5\; , \quad
\Delta\eta_{\ell j}={\rm min}(|\eta_{\ell}-\eta_{j_1}|,
|\eta_{\ell}-\eta_{j_2}|) > 2.5\; ,
\end{equation}
and already leads to a signal slightly larger than the QCD background
(620~fb vs. 500~fb).

A further strong background rejection is achieved by exploiting the very
large dijet invariant masses which are typical for the vector boson
fusion process. The $m_{jj}$-distributions (after imposing the cuts of
Eq.~\ref{separationlepton}) are shown in Fig.~3.
\vglue3.0in
\includegraphics{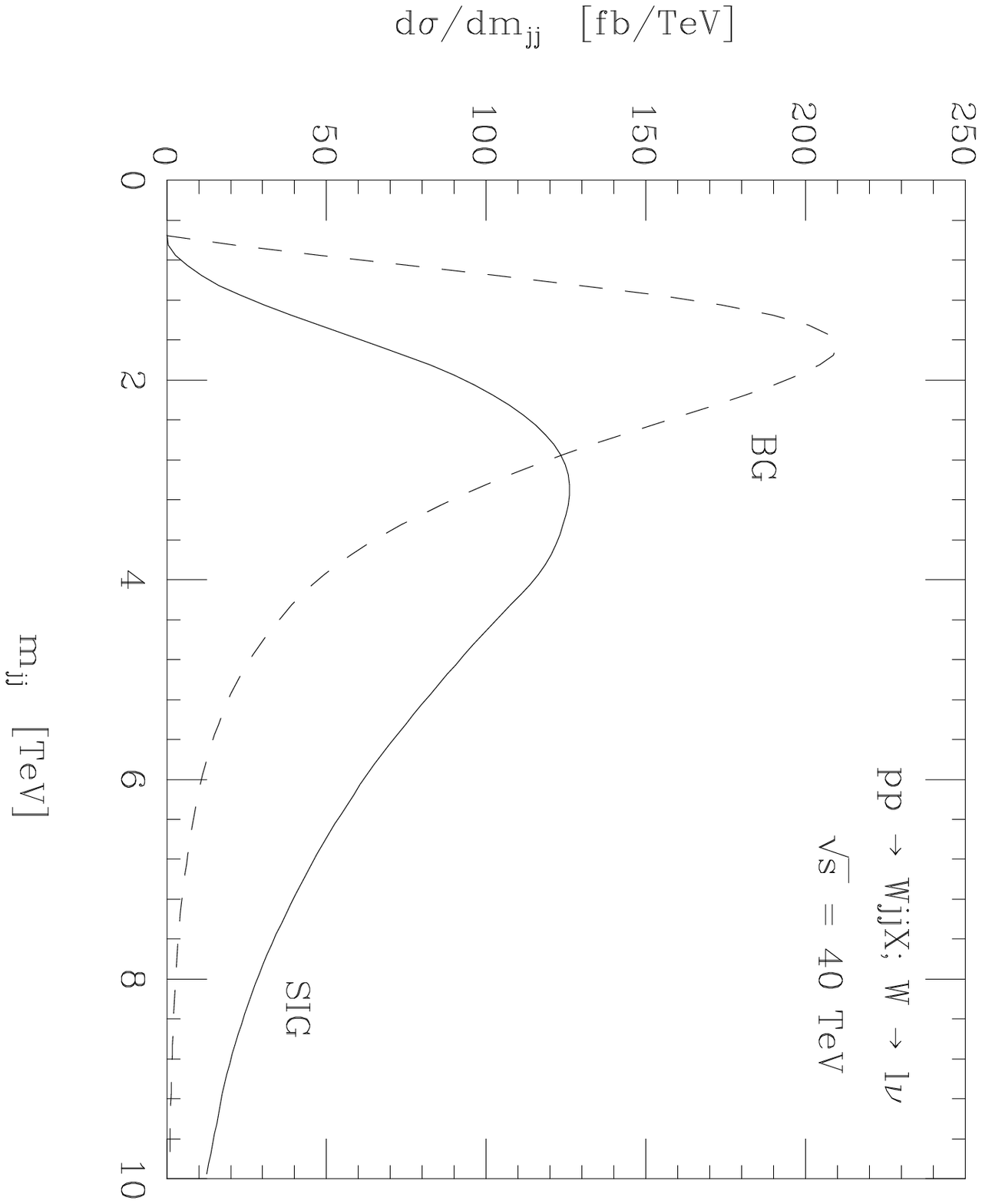}
\begin{center}
\parbox{13cm}{\small\baselineskip12pt
 Fig.~3: Invariant mass distribution of the two tagging jets for the signal
(solid line) and the QCD background (dashed line) at the SSC. }
\end{center}

A dijet invariant mass cut of
\begin{equation}\label{mjjcut}
m_{jj}>3~{\rm TeV},
\end{equation}
imposed on the two tagging jets, reduces the background well below signal
level
($\sigma_{\rm SIG} = 450$~fb vs. $\sigma_{\rm QCD} = 136$~fb for the QCD $Wjj$
background and $\sigma_{t\bar tj} = 38$~fb for the top-quark background at
$m_t=140$~GeV).

\section{Sensitivity to the $WWV$ Vertex}

The cuts discussed in the previous section single out the phase space region
in which the electroweak fusion process dominates and hence we expect a
pronounced sensitivity to deviations in the TGV's from the SM prediction.
Because of the extra derivatives in the operators of the effective Lagrangian
of Eq.~\ref{LeffWWV} and the now incomplete gauge theory cancelations for
longitudinal polarization of the incoming gauge bosons, anomalous coupling
effects are enhanced at large momentum transfer and, hence, for large
transverse momenta of the produced $W$-boson.  The effect is demonstrated in
Fig.~4.
\begin{figure}[t]
\vglue3.4in
\includegraphics{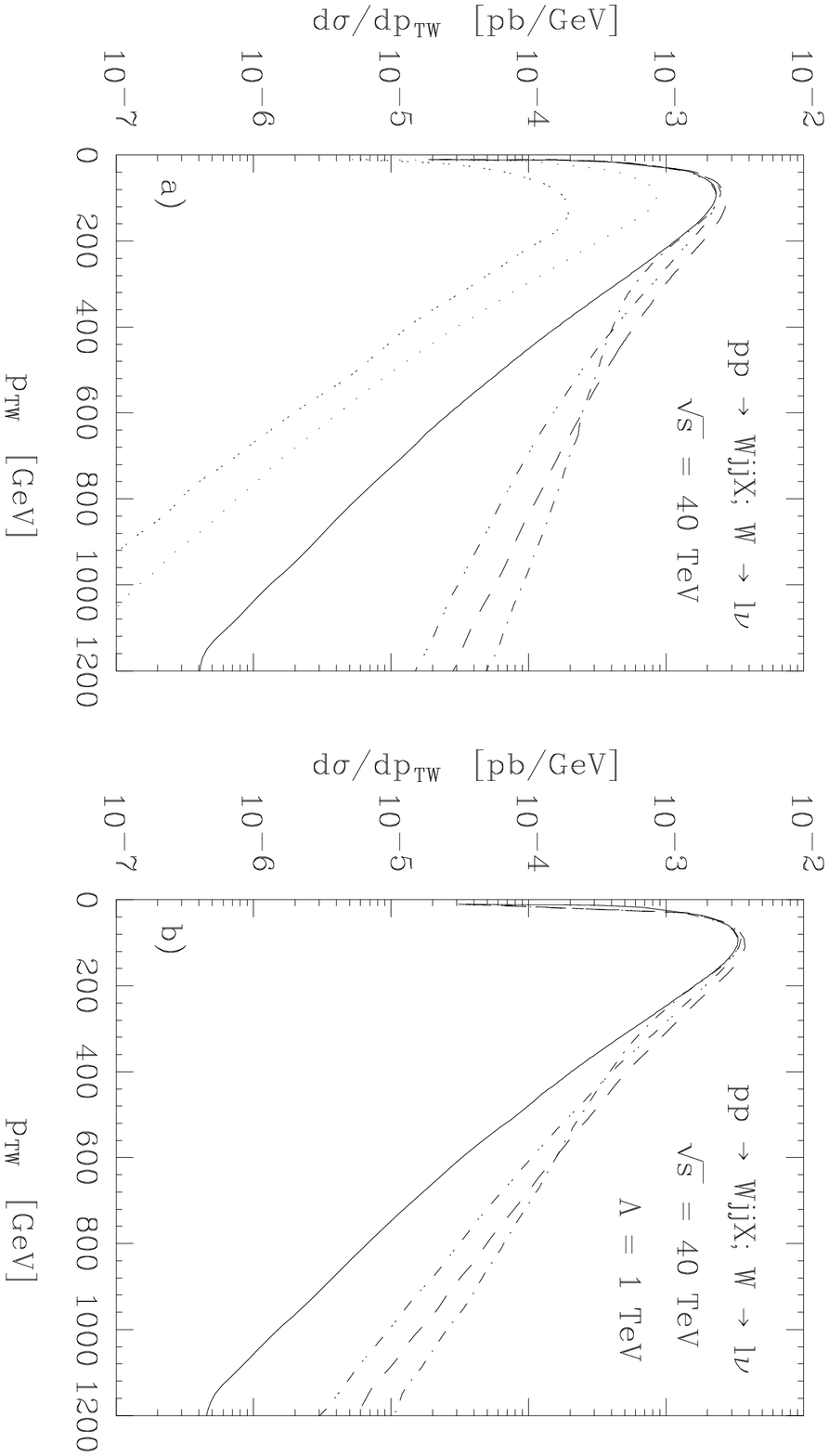}
\begin{center}
\parbox{13cm}{\small\baselineskip12pt
 Fig.~4: Transverse momentum distribution of the produced $W$-boson in
$Wjj$ events at the SSC. Part a) shows individual distributions for the SM
signal (solid line) the QCD $Wjj$ background (dotted line) and the $t\bar
tj$ background for $m_t=140$~GeV (double dotted line). The upper three
curves correspond to three choices of anomalous couplings: $\kappa_\gamma=
\kappa_Z=1.2$ (dashed line) $\lambda_\gamma=\lambda_Z=0.1$ (dash-dotted curve)
and $g_1^Z=1.2$ (dash-double dotted line). In part b) the two background
distributions have been added to the four signal curves. In addition the
effect of the form-factor of Eq.~\ref{formf} is shown for a scale
$\Lambda=1$~TeV. The cuts imposed are described in the text.}
\end{center}
\vskip -0.5cm
\end{figure}
While the $p_{TW}$ distributions show similar shapes for the SM
signal and the QCD and top quark backgrounds, a strong enhancement at large
transverse momenta arises from anomalous couplings like $\kappa_\gamma-1=
\kappa_Z-1=0.2$ (dashed line) $\lambda_\gamma=\lambda_Z=0.1$ (dash-dotted
curve) and $g_1^Z-1=0.2$ (dash-double dotted line). For these three curves all
other anomalous couplings are set to zero. We actually find the $p_{TW}$
distribution to be the one which is the single most sensitive to anomalous
TGV's. The dotted and double dotted curves in Fig.~4a represent the QCD
$Wjj$ and the $t\bar tj$ background (with $m_t=140$~GeV), respectively.
For a smaller top quark mass of $m_t=110$~GeV, the $p_{TW}$
differential cross section
from $t\bar tj$ production is enhanced by roughly a factor two in the
peak region around $p_{TW}\approx 100$~GeV. For $W$ transverse momenta larger
than 400~GeV, where deviations from the SM are most pronounced, there
is only little variation of the differential cross section with $m_t$.

The anomalous couplings used in Fig.~4a are of the same order or below
the sensitivity limits expected for $W^+W^-$ production at LEP
II~\cite{LEPWW}.
However, at LEP II these couplings are probed at relatively small time-like
momentum transfers of order $q^2 = (200~{\rm GeV})^2$, while the single $W$
production process is most sensitive at large space-like momentum transfers
of order $-q^2 \approx p_{TW}^2 = (1~{\rm TeV})^2$. These different scales
raise the possibility that form-factor effects may be important in the
$Wjj$ production
considered here. We have shown previously\cite{BZ} that anomalous
couplings as large as the ones used in Fig.~4 do require form-factor damping
for a momentum transfer above $|q|=2$~TeV in order not to violate partial
wave unitarity in vector boson pair production.
By analytic continuation we expect a similar form-factor scale also in the
space-like region.

In Fig.~4b we show results including such form-factor effects. We have
replaced the anomalous TGV's $a=\kappa_V-1,\; g_1^Z-1$, or $\lambda_V$ by
\begin{equation}\label{formf}
a\to {a\over (1+ |q_1^2|/ \Lambda^2) (1+|q_2^2|/ \Lambda^2)
(1+|q_3^2|/ \Lambda^2) }\; ,
\end{equation}
with a cutoff scale $\Lambda=1$~TeV. Here $q_1$, $q_2$, and $q_3$ denote the
four-vectors of the three vector bosons entering the TGV in Fig.~1.

Form-factor effects are clearly important in the measurement of TGV's at the
SSC; the very large enhancements above $p_{TW}=1$~TeV in Fig.~4a are most
likely unrealistic and the situation in Fig.~4b represents a more likely
scenario. Nevertheless, $qq\to qqW$ production is a sensitive probe for
anomalous couplings. In the region above $p_{TW}=400$~GeV the anomalous
coupling curves in Fig.~4b correspond to an expected rate of 600--800
$W\to e\nu,\; \mu\nu$ events per SSC year (10 fb$^{-1}$) as compared to about
200 events for the SM.

The results of a more quantitative analysis of the sensitivity of $Wjj$
production to anomalous TGV's are shown in Table~1 where we list the
$2\sigma$ limits achievable for $\Lambda=1$~TeV at the
SSC. Only one coupling at a time is assumed to differ from its SM value.
In order to demonstrate the effect of the form-factor behavior of the
anomalous couplings, we also display sensitivity bounds for
$\Lambda=\infty$, which is equivalent to ignoring the momentum
dependence of the non-standard $WWV$ couplings.
\begin{table}[t]
\centering
\caption
[limits.]
{Sensitivities achievable at the $2\sigma$ level for the anomalous
$WWV$, $V=\gamma,\, Z$ couplings $\Delta\kappa_V=\kappa_V-1$,
$\lambda_V$, and $\Delta g_1^Z=g_1^Z-1$ in $pp\rightarrow W^\pm
jj\rightarrow\ell^\pm\nu jj$ at the SSC. Only one coupling at a time is
varied. We assume an integrated luminosity of 10~fb$^{-1}$. The cuts
imposed and the form-factor ansatz used are described in the text. The
limits are displayed for form-factor scales of $\Lambda=1$~TeV and
$\Lambda=\infty$.
}
\vspace{6.mm}
\begin{tabular}{|c|c|c|c|c|c|}\hline
coupling & $\Lambda=1$~TeV & $\Lambda=\infty$ & coupling &
$\Lambda=1$~TeV & $\Lambda=\infty$ \\ \hline
$\Delta\kappa_\gamma$ & $\matrix{+0.09 \crcr\noalign{\vskip -3pt} -0.
18}$ & $\matrix{+0.06 \crcr\noalign{\vskip -3pt} -0.11}$ &
$\Delta\kappa_Z$ & $\matrix{+0.05 \crcr\noalign{\vskip -3pt} -0.09}$ &
$\matrix{+0.04 \crcr\noalign{\vskip -3pt} -0.05}$ \\[2.mm]
$\lambda_\gamma$ & $\matrix{+0.05 \crcr\noalign{\vskip -3pt} -0.05}$ &
$\matrix{+0.03 \crcr\noalign{\vskip -3pt} -0.03}$ &
$\lambda_Z$ & $\matrix{+0.03 \crcr\noalign{\vskip -3pt} -0.03}$ &
$\matrix{+0.02 \crcr\noalign{\vskip -3pt} -0.02}$ \\[2.mm]
$\Delta g_1^\gamma$ & -- & -- &
$\Delta g_1^Z$ & $\matrix{+0.05 \crcr\noalign{\vskip -3pt} -0.11}$ &
$\matrix{+0.03 \crcr\noalign{\vskip -3pt} -0.06}$ \\[2.mm] \hline
\end{tabular}
\label{tab:xsect}
\end{table}
We calculate the statistical significance by splitting the $p_{TW}$
distribution into 12 bins, 11 of which are 60 GeV wide.
In each bin the Poisson statistics is
approximated by a Gaussian distribution. In order to achieve a sizable
counting rate in each bin, all events with $p_{TW}>660$~GeV are
collected in one bin. To derive realistic limits we allow for a
normalization uncertainty of 50\% in the SM cross section. The QCD $Wjj$
and $t\bar tj$ background contributions are fully incorporated in our
procedure.

The results collected in Table~1 indicate that the $WWV$ vertices can be
probed in $qq\rightarrow qqW$ at the few percent level in general. Form-factor
effects may weaken the achievable bounds by up to a factor 1.5, for
form-factor scales above 1~TeV.
Comparing our results with those obtained in Ref.~12, we find that the process
$qq\rightarrow qqW$ is significantly more sensitive to $\Delta\kappa_V$
and $\Delta g_1^Z$ than $W\gamma$ and $WZ$ production for cutoff scales
$\Lambda$ in the low TeV range. The measurement of $\lambda_V$ is competitive.
In general, the pair production process is affected more by details of the
form-factors, which in addition may be quite different in the space-like and
the time-like regions. This emphasizes the need to measure pair production and
$Wjj$ production if full information on the $WWV$ couplings is to
be gained.

\section*{Acknowledgements}
This research was supported in part by the University of Wisconsin Research
Committee with funds granted by the Wisconsin Alumni Research Foundation,
by the U.~S.~Department of Energy under contracts No.~DE-AC02-76ER00881
and DE-FG05-87ER40319, and
by the Texas National Research Laboratory Commission under Grants
No.~RGFY9273 and FCFY9212.

\end{document}